\documentclass[sigconf]{acmart}
\AtBeginDocument{%
  }

\copyrightyear{2026}
\acmYear{2026}
\setcopyright{cc}
\setcctype{by-nc-nd}
\acmConference[RecSys '26]{20th ACM Conference on Recommender Systems}{September 27-October 02, 2026}{Minneapolis, MN, USA}
\acmBooktitle{20th ACM Conference on Recommender Systems (RecSys '26), September 27-October 02, 2026, Minneapolis, MN, USA}
\acmDOI{10.1145/3773078.3831906}
\acmISBN{979-8-4007-2284-4/2026/09}

\usepackage{amsmath}
\usepackage{booktabs}
\usepackage{graphicx}
\usepackage{microtype}
\usepackage{subfigure}

\begin{document}

\title{LO-FAR: A Cost-Aware Local Filter for Sparse Feature Ranking in Industrial Ad Recommendation}

\author{Egemen Erbayat}
\email{egemen@meta.com}
\affiliation{%
  \institution{Meta Platforms, Inc.}
  \country{USA}
}\authornote{This paper was completed during Egemen Erbayat's internship at Meta Platforms, Inc..}

\author{Luis Duque}
\email{luisduque@meta.com}
\affiliation{%
  \institution{Meta Platforms, Inc.}
  \country{USA}
}

\author{Sohini Roychowdhury}
\email{sroychowdhury1@meta.com}
\affiliation{%
  \institution{Meta Platforms, Inc.}
  \country{USA}
}

\author{Mohammad Amin}
\email{shafkat@meta.com}
\affiliation{%
  \institution{Meta Platforms, Inc.}
  \country{USA}
}

\author{Srihari Reddy}
\email{sriharir@meta.com}
\affiliation{%
  \institution{Meta Platforms, Inc.}
  \country{USA}
}

\renewcommand{\shortauthors}{Erbayat et al.}
\acmArticleType{Review}
\keywords{Recommender systems; Online advertising; Sparse ID-list features; Feature selection; CTR/CVR prediction; Embedding tables; Industrial machine learning; Normalized Entropy.}
\begin{abstract}
Industrial ad recommendation models rely heavily on sparse, high-cardinality ID-list features that encode user histories and contextual identifiers. Each is backed by a dedicated embedding table, so these features dominate storage, training, and serving cost and must be revisited as traffic and downstream models evolve. Therefore, sparse feature ranking is not just an offline modeling problem but also a recurrent systems decision limited by compute budgets and iteration cadence. We present Localized Feature Ranking (LO-FAR), a CPU-only, model-agnostic workflow that ranks each candidate feature from its stand-alone held-out predictive signal using lightweight local estimators rather than the GPU-bound retraining loops of permutation- and stochastic-gate-based methods. On a production dataset of more than one million logged interactions and 475 sparse ID-list features, LO-FAR completes ranking in approximately two CPU-hours and preserves downstream Normalized Entropy gains on CTR and CVR tasks that are competitive with shuffle-based importance, Binary Stochastic Neurons, and a coverage-based heuristic across budgets of 100--400 retained features. The contribution is a deployable workflow showing that, when cost and turnaround constraints are binding, a simple local filter can be a practical production choice over heavier interaction-aware alternatives.
\end{abstract}

\maketitle

\section{Introduction}\label{sec:intro}

Industrial ad recommendation systems are large-scale deployed pipelines that must score thousands of candidate items per request under tight latency budgets, while continuously adapting to shifting user behavior, advertiser inventory, and platform policies \cite{richardson2007predicting,he2014practical,cheng2016wide,covington2016deep,naumov2019deeplearningrecommendationmodel}. The core supervised learning problem in these pipelines is to predict the probability that a given (user, context, ad) triple will result in a desired outcome, most commonly a click in click-through rate (CTR) prediction or a downstream action in conversion rate (CVR) prediction. Deep models that mix dense numerical features with {sparse ID-list features}, which are variable-length lists of high-cardinality categorical IDs that encode entities like user histories, page elements, ad identifiers, and context tokens, are used by modern rankers to address these issues\cite{naumov2019deeplearningrecommendationmodel,zhao2023embedding}. Each sparse feature is associated with a dedicated embedding table that maps its identifier vocabulary into a dense vector space.

On the other hand, sparse ID-list features dominate the parameter count, memory footprint, and infrastructure cost of contemporary rankers. In the production-grade model studied in this work, sparse embedding tables account for over 97\% of all trainable parameters; similar proportions have been reported across industrial systems \cite{liu2022monolith,kang2021dhe,li2024embeddingcompression}. Because each feature has its own vocabulary, lookup path, monitoring contract, and refresh schedule, every additional sparse feature adds engineering surface area beyond its training-time cost. Consequently, the question of {which} sparse features to keep is not only a modeling decision but an operational decision that shapes storage, serving memory, retraining throughput, and the velocity of downstream feature work.

At the same time, sparse feature ranking is not a one-shot offline study in an industrial setting. It must be repeated whenever new candidate features are introduced, vocabularies drift, traffic distributions shift, or downstream model architectures change \cite{lyu2025dv365,yin2024automl}. Each rerun consumes engineering time, accelerator capacity, and review effort. A ranking method whose offline accuracy is marginally higher but whose rerun cost is substantially larger may therefore yield a worse system-level outcome, because pruning decisions become stale before they can be re-validated. The dominant academic baselines for high-quality sparse feature selection illustrate this trade-off. Permutation-based importance \cite{fisher2019all} requires repeated full-model retraining cycles, one per feature or feature group, on accelerator hardware. Stochastic-gate methods, such as those built on $L_0$ regularization and Binary Stochastic Neurons (BSN) \cite{louizos2018learningsparseneuralnetworks,yamada2020feature}, embed selection into end-to-end training and inherit its computational profile. Recent recommender-system work on training-coupled feature and embedding optimization has further sharpened this coupling \cite{lyu2022optembed,lyu2023optfs,lin2022adafs}. These methods can recover interaction-sensitive signal that simpler procedures cannot, but they are difficult to operate at the cadence required by production teams.

We address a complementary operating point. Specifically, we focus on {first-stage} pruning of large sparse feature pools under explicit storage and retraining budgets, where the value of a method is determined by the joint quality, turnaround time, and operational simplicity of repeated reruns. We propose Localized Feature Ranking (LO-FAR), a CPU-only ranking workflow that scores each candidate sparse ID-list feature from its stand-alone held-out predictive signal using lightweight local estimators, without invoking the downstream ranker during the ranking pass. LO-FAR processes each feature independently, parallelizes linearly across CPU workers, and produces a ranked list that downstream training pipelines can consume directly. We position LO-FAR as a first-stage filter rather than a replacement for interaction-aware methods. It is intended to remove the long tail of low-signal sparse features cheaply, which leaves more expensive procedures available for cases where they are clearly justified.

We evaluate LO-FAR on a production-grade dataset comprising more than one million logged interactions and 475 short sparse ID-list features drawn from a large ad recommendation system. We compare against three baselines selected to span the operational spectrum: a coverage-based naive heuristic, shuffle-based permutation importance \cite{fisher2019all}, and BSN \cite{louizos2018learningsparseneuralnetworks,yamada2020feature}. Following standard practice in industrial CTR/CVR modeling, we evaluate downstream ranker quality with Normalized Entropy (NE) gain relative to a dense-only reference model trained without sparse features \cite{he2014practical}, measured at feature budgets of 100, 200, 300, and 400 retained features. Across both CTR and CVR tasks and all four budgets, LO-FAR preserves NE gains that are comparable to or better than shuffle-based importance and BSN, while completing ranking in approximately two CPU-hours. In the same environment, the heavier baselines required multiple GPU-days when accumulated retraining and tuning runs are accounted for. Because sparse embeddings dominate parameter count in our model, the chosen feature budgets correspond to deterministic reductions of approximately 40--75\% in sparse embedding storage.

Taken together, this paper makes the following contributions:
\begin{enumerate}
    \item We frame industrial sparse ID-list feature ranking as a recurring operational problem in which rerun cost, accelerator dependency, and integration with existing training infrastructure are first-class evaluation criteria, and we formalize the recurring-cost view in Section~\ref{sec:methodology}.
    \item We present LO-FAR, a CPU-only, model-agnostic local ranking workflow that scores each sparse ID-list feature from its stand-alone held-out predictive signal, provide its computational complexity, and characterize its scope and limitations explicitly.
    \item We report a production-scale evaluation on more than 1M logged interactions and 475 sparse ID-list features. LO-FAR achieves downstream NE gains competitive with shuffle-based importance and BSN across feature budgets of 100--400 retained features and completes ranking in approximately two CPU-hours.
    \item We describe the deployment implications, including 40--75\% deterministic reductions in sparse embedding storage at the studied budgets, the operational comparison against shuffle-based importance and BSN, and the conditions under which LO-FAR should not be used in isolation.
\end{enumerate}

\section{Related Work}\label{sec:related_work}

We organize related work along four axes that frame the contribution of this paper. These include industrial CTR/CVR ranking models that depend on large sparse feature spaces, interaction-learning architectures that intensify the value and the cost of sparse capacity, training-coupled feature and embedding selection methods, and model-agnostic importance estimation. Table~\ref{tab:lit_schools} summarizes these directions and their relation to LO-FAR.

\subsection{Industrial CTR/CVR Models and Sparse Feature Economics}

Modern industrial recommenders evolved from logistic regression and tree-augmented linear models for CTR prediction \cite{richardson2007predicting,he2014practical} into deep architectures that combine memorization and generalization through embedded categorical features and dense numerical inputs \cite{cheng2016wide,covington2016deep,naumov2019deeplearningrecommendationmodel}. Across this trajectory, sparse categorical features have remained the dominant carrier of personalization signal. Recent work extends this tradition along two distinct lines. The first emphasizes capacity scaling and richer interaction modeling for web-scale ranking, exemplified by DCN~V2 \cite{wang2021dcnv2}, large-capacity AutoML search at Meta \cite{yin2024automl}, and long-history user representation at Instagram \cite{lyu2025dv365}. The second confronts the systems burden imposed by sparse embeddings through hash-based table alternatives \cite{kang2021dhe,liu2021pep}, distributed training systems for huge embedding tables \cite{guo2021scalefreectr,xu2021agilectr,sethi2022recshard}, online collisionless embedding infrastructure \cite{liu2022monolith}, and surveys of embedding compression techniques \cite{li2024embeddingcompression}. Recent work on cost-effective model architecture design at industry scale~\cite{luo2025metalattice} further underscores the need for principled decisions about which features and model components to retain under budget constraints. These two directions share an underlying premise. Sparse features carry substantial behavioral signal, but every additional feature imposes recurring cost in the form of vocabulary management, embedding-table lookups, monitoring, and retraining. The literature does not resolve this trade-off; it characterizes the engineering required to sustain it.

\subsection{Interaction Learning and Adaptive Sparse Capacity}

A complementary line of work targets the explicit modeling of feature interactions in sparse spaces. Factorization-machine--based architectures, such as DeepFM and its variants \cite{guo2017deepfm,10.1145/3671151.3671161}, learn pairwise interactions in low-dimensional embedding spaces. Automatic interaction selection methods, such as AutoFIS, prune the resulting interaction lattice during training \cite{liu2020autofis}. Adaptive sparse architectures, including AdaSparse and AdaFS, allocate model capacity dynamically across domains and tasks \cite{yang2022adasparse,lin2022adafs}. Sequential and transformer-based architectures, such as SASRec and Hiformer, broaden the effective sparse context available to the ranker \cite{kang2018self,gui2023hiformer}. While these methods raise the achievable quality ceiling, they also intensify the operational problem. As interaction depth and history length increase, the marginal cost of each additional candidate sparse feature grows in tandem. This makes principled, low-cost first-stage pruning more rather than less important.

\subsection{Training-Coupled Feature and Embedding Selection}

Several recent methods integrate feature selection into model training. Stochastic-gate methods learn binary or relaxed gates over input features under $L_0$ regularization, with BSN as a representative instance \cite{louizos2018learningsparseneuralnetworks,yamada2020feature}. OptEmbed and OptFS jointly optimize embedding structure and feature subsets within the training loop \cite{lyu2022optembed,lyu2023optfs}. AutoField~\cite{wang2022autofield} further automates feature-level selection for deep recommender systems through a controller network that learns to include or exclude input features during training. The strength of this family is that selection is performed in the presence of feature interactions and correlations, so that features whose value emerges only in combination can in principle be retained. The corresponding limitation is operational. Ranking decisions are tied to the downstream model architecture, the training hyperparameters, and the availability of accelerator-heavy retraining. Each change in the ranker, the candidate feature pool, or the training data may require rerunning the joint optimization. In high-churn industrial settings, where heterogeneity in user behavior is the norm rather than the exception \cite{10.1145/3485447.3511950}, and where infrastructure papers devote substantial engineering effort to keeping the training loop fast and stable under load \cite{guo2021scalefreectr,xu2021agilectr}, this coupling becomes a non-trivial cost.

\subsection{Model-Agnostic Importance Estimation}

Model-agnostic methods evaluate the importance of input features after training without modifying the model itself. Permutation importance \cite{fisher2019all} and its variants estimate the marginal contribution of each feature by measuring the degradation of model performance under random permutation of feature values. These methods are conceptually simple and architecturally portable, but they require repeated full-model evaluation passes and, in CTR/CVR settings with large embedding tables, repeated retraining or recalibration to obtain stable estimates. Statistical filter methods such as mutual information offer linear scaling in the number of features but were developed for low-cardinality or moderately sparse regimes \cite{fan2008sure,krishnan2023mutual,peng2005feature}. Wrapper and metaheuristic approaches \cite{kohavi1997wrappers,xue2015survey} provide accuracy guarantees on small problems but do not scale to feature pools that contain hundreds to thousands of high-cardinality sparse ID-list features. The practical consequence is that, despite a rich literature on importance estimation, there is no widely adopted procedure that simultaneously satisfies the four requirements relevant to industrial first-stage pruning. These requirements include high feature count, embedding-aware treatment of sparse ID-list inputs, low and predictable turnaround on standard infrastructure, and decoupling from the downstream training loop.

Concurrent work by Jia et al.~\cite{jia2024erase} provides a comprehensive benchmark of feature selection methods for deep recommender systems, evaluating gate-based, gradient-based, and sensitivity-based approaches. LO-FAR differs by targeting the industrial operating point of short sparse ID-list features under explicit cost and turnaround constraints, with evaluation through end-to-end retraining rather than proxy metrics alone.

\subsection{Positioning of LO-FAR}

LO-FAR addresses the gap identified in the preceding subsections. It is not positioned against the broader literature on representation learning or interaction modeling; rather, it complements that literature by making the first-stage decision of {which} sparse ID-list features to retain cheap, repeatable, and operable on standard CPU infrastructure. We frame this contribution as a system-level reformulation. Importance estimation for sparse ID-list features is decoupled from full-model retraining and, instead, is reduced to a per-feature held-out prediction problem solved by a lightweight local estimator. We emphasize that ranking divergence between LO-FAR and interaction-aware methods is not a defect but a structural property. In correlated sparse spaces, multiple feature subsets may encode overlapping signal, so two ranking methods can disagree while still inducing comparable downstream model quality once the retained subsets are retrained.

\begin{table*}[t]
\centering
\footnotesize
\setlength{\tabcolsep}{3pt}
\caption{Schools of thought in recent industrial recommendation literature and their relation to LO-FAR. The first three rows describe complementary directions; LO-FAR addresses a distinct operating point that those directions do not target.}
\label{tab:lit_schools}
\begin{tabular}{p{0.18\textwidth} p{0.31\textwidth} p{0.20\textwidth} p{0.23\textwidth}}
\toprule
School of thought & Representative recent work & Dominant objective & Relation to our setting \\
\midrule
Capacity scaling and richer interactions & DCN~V2 \cite{wang2021dcnv2}, AdaSparse \cite{yang2022adasparse}, AdaFS \cite{lin2022adafs}, DV365 \cite{lyu2025dv365}, Meta AutoML \cite{yin2024automl} & Increase ranking quality by expanding expressive capacity and modeling higher-order crosses & Assumes that firms can sustain repeated optimization of increasingly complex sparse stacks \\
Embedding efficiency and infrastructure optimization & DHE \cite{kang2021dhe}, ScaleFreeCTR \cite{guo2021scalefreectr}, agile CTR training \cite{xu2021agilectr}, Monolith \cite{liu2022monolith}, and the embedding-compression survey \cite{li2024embeddingcompression} & Keep large sparse spaces trainable and serveable through systems-level engineering & Reduces the cost of retaining many sparse features but does not decide which features to retain \\
Training-coupled feature and embedding selection & BSN-style gates \cite{louizos2018learningsparseneuralnetworks,yamada2020feature}, OptEmbed \cite{lyu2022optembed}, OptFS \cite{lyu2023optfs} & Select features and embedding structure jointly with the downstream model & Captures interaction-sensitive signal but couples ranking cost to the training stack \\
First-stage operational filtering (this paper) & LO-FAR & Reduce the sparse candidate pool before expensive retraining under explicit budgets & Sacrifices interaction awareness for cheap reruns, linear CPU parallelization, and predictable cadence \\
\bottomrule
\end{tabular}
\end{table*}

\section{Production Setting and Problem Definition}\label{sec:methodology}

\subsection{Deployed Ranking Model}\label{subsec:dlrm}

The system studied in this work is a deployed industrial ad ranker built on the Deep Learning Recommendation Model (DLRM) family of architectures \cite{naumov2019deeplearningrecommendationmodel,acun2021understanding}, trained on logged user--ad interactions and used to produce CTR and CVR predictions. The model consumes three categories of input:
\begin{itemize}
    \item \textbf{Binary features} encode true/false events, such as whether a previous user-level signal is present.
    \item \textbf{Dense features} are continuous numerical inputs, such as engagement durations and aggregate counters, processed by a multilayer perceptron (MLP).
    \item \textbf{Sparse ID-list features} are variable-length lists of high-cardinality categorical identifiers, such as recently seen page or ad IDs. Each such feature is associated with a dedicated embedding table that maps its identifier vocabulary to dense vectors. Variable-length lists are pooled (typically by sum or mean) before being combined with dense and binary representations.
\end{itemize}
The dense, binary, and pooled-sparse representations are fused through MLP transformations and pairwise dot-product feature interactions \cite{naumov2019deeplearningrecommendationmodel}, and the resulting representation is mapped to a probability through a logistic head. In the deployed model studied here, sparse embedding tables account for more than 97\% of all trainable parameters and are the primary driver of storage, training memory, and serving footprint.

\subsection{Notation and Problem Statement}

Let $\mathcal{D} = \{(x_i, y_i)\}_{i=1}^{n}$ denote a logged training sample, where $x_i$ is the feature vector for the $i$-th interaction and $y_i \in \{0, 1\}$ is a binary label encoding the supervised target (a click for CTR; a downstream action for CVR). Let $\mathcal{F} = \{f_1, \ldots, f_p\}$ denote the candidate set of $p$ sparse ID-list features. For each feature $f_j$ and example $i$, the feature value is an unordered list of categorical identifiers
\begin{equation}
x_i^{(j)} = \bigl[\,id_{i,1}^{(j)},\, id_{i,2}^{(j)},\, \ldots,\, id_{i,k_i}^{(j)}\,\bigr],
\end{equation}
where $k_i$ is the (variable) list length for example $i$ and feature $j$, and the identifier $id_{i,t}^{(j)}$ is drawn from the vocabulary of feature $f_j$. Empty lists are admissible and are filled with a placeholder identifier.

The objective of sparse feature ranking is to produce an ordering $\pi: \{1,\ldots,p\} \to \{1,\ldots,p\}$ over $\mathcal{F}$ such that, for any practitioner-chosen budget $B \le p$, the top-$B$ subset $\mathcal{F}_B = \{f_{\pi(1)}, \ldots, f_{\pi(B)}\}$ supports a downstream ranker whose end-to-end quality is competitive with the model trained on the full sparse pool. The evaluation criterion is downstream NE gain (Section~\ref{subsec:ne}) after retraining the ranker with the selected subset, not agreement with any other ranking method.

\subsection{Operational Requirements}

In industrial pipelines, sparse feature ranking is rerun whenever new candidate features arrive, vocabularies are refreshed, traffic distributions shift, or the downstream model is updated. A ranking workflow that is to be useful in this regime must satisfy four practical requirements:
\begin{enumerate}
    \item it must scale to hundreds to thousands of sparse features without quadratic cost in $p$;
    \item it must have predictable turnaround on commodity infrastructure, ideally without contending with accelerator capacity needed for model training and large-scale evaluation;
    \item it must be simple to operate, parallelize, monitor, and rerun within existing data pipelines; and
    \item it must preserve enough downstream model quality at the chosen feature budget to justify the pruning decision.
\end{enumerate}
LO-FAR is designed for this operating point. It does not attempt to solve full joint feature selection for the downstream ranker. Instead, it provides a low-cost first-stage filter that removes low-signal sparse features before more expensive procedures are invoked.

\subsection{Storage Implications of Pruning}

The reported storage reduction is a deterministic consequence of pruning, not a separate empirical metric. Each retained sparse ID-list feature $f_j$ contributes an embedding table whose size is determined by its vocabulary cardinality and embedding dimension. Removing $f_j$ removes the corresponding table from both training and serving artifacts. In the studied model, sparse tables account for more than 97\% of the parameter count, so retaining $B$ out of $p = 475$ features removes approximately $1 - B/p$ of the sparse parameter budget under the (approximately satisfied) assumption that retained and removed features have comparable per-table size. For $B \in \{100, 200, 300, 400\}$, this corresponds to roughly $79\%$, $58\%$, $37\%$, and $16\%$ removal, respectively, which we report as the practitioner-relevant range of 40--75\% sparse-storage reduction at the budgets that preserve downstream NE most effectively.

\subsection{A Recurring-Cost View of Feature Ranking}\label{subsec:recurring_cost}

The choice of ranking workflow has consequences that compound over time. To make this explicit, consider a workflow $m$, a feature budget $B$, and a refresh cadence $R$ (the number of times the workflow must be rerun per accounting period). The recurring cost incurred by the firm can be decomposed as
\begin{equation}
\begin{aligned}
C(m, B, R) &= C_{\text{rank}}(m,\, R) \\
&\quad + R \cdot \bigl( C_{\text{train}}(B) + C_{\text{store}}(B) + C_{\text{serve}}(B) \bigr),
\end{aligned}
\label{eq:recurring_cost}
\end{equation}
where $C_{\text{rank}}$ aggregates the cost of running the ranking workflow itself, and $C_{\text{train}}$, $C_{\text{store}}$, and $C_{\text{serve}}$ are the per-cycle training, storage, and serving costs at the chosen budget $B$. Equation~\eqref{eq:recurring_cost} is not a complete cost model; its purpose is to make explicit a structural fact often left implicit in the literature. Training-coupled selection methods reduce $C_{\text{train}}$, $C_{\text{store}}$, and $C_{\text{serve}}$ at the chosen budget but increase $C_{\text{rank}}$, because importance estimation and downstream training are entangled. A first-stage local filter targets the same downstream savings while keeping $C_{\text{rank}}$ small and predictable, which becomes important when $R$ is large. In ad recommendation systems with multiple ranking tasks, several feature owners, and frequent retraining schedules, even modest reductions in $C_{\text{rank}}$ can compound into substantial savings, and the marginal cost of a sparse feature must be evaluated relative to its recurring contribution to all four terms in equation~\eqref{eq:recurring_cost}. The challenge of feature freshness in production CTR systems has been studied from the model update perspective~\cite{fesail2023}, where stale feature embeddings degrade prediction quality over time; this reinforces the argument that ranking workflows must be rerun frequently and cheaply.

\section{LO-FAR: A Local Ranking Workflow for Sparse ID-List Features}
\label{sec:local_ranking}

LO-FAR ranks each candidate sparse ID-list feature $f_j \in \mathcal{F}$ from its stand-alone held-out predictive signal, computed by a lightweight local estimator that operates only on the values of $f_j$ and the supervised label $y$. This per-feature independence is the central design choice. It eliminates the need to invoke or retrain the downstream ranker during the ranking pass, makes parallelization across features trivial, and reduces the ranking workflow to a CPU-only batch job that fits standard data pipelines.

\subsection{Workflow}

LO-FAR consists of six stages applied uniformly to each candidate sparse feature:
\begin{enumerate}
    \item \textbf{Sampling.} Draw a representative subset of the logged data (in our experiments, $n \approx 10^6$ rows) to balance statistical fidelity against ranking turnaround time. Stratified sampling by the supervised label is used to preserve the empirical positive rate.
    \item \textbf{Train/test split.} Partition the sampled rows into $\mathcal{D}_{\text{train}}$ and $\mathcal{D}_{\text{test}}$ to enable held-out evaluation of feature-specific predictors. Splits are stratified by label.
    \item \textbf{Feature explosion.} For each example $i$, the variable-length list $x_i^{(j)} = [id_{i,1}^{(j)}, \ldots, id_{i,k_i}^{(j)}]$ is unrolled into one row per identifier, with the example label $y_i$ replicated across rows. Empty lists are filled with a placeholder identifier $-1$. Formally,
    \begin{equation}\label{eq:explosion}
    \mathcal{D}^{(j)}_{\text{train,exp}} = \bigl\{(id_{i,t}^{(j)},\, y_i)\,:\, t = 1,\ldots,k_i,\, i \in I_{\text{train}}\bigr\},
    \end{equation}
    and analogously for $\mathcal{D}^{(j)}_{\text{test,exp}}$.
    \item \textbf{Local estimator.} A lightweight ID-level predictor $s_j: \text{Vocab}(f_j)$ $\to [0, 1]$ is fit on $\mathcal{D}^{(j)}_{\text{train,exp}}$. For an identifier $id$ that occurs at least $K$ times in the exploded training data, $s_j(id)$ is set to the empirical positive rate of that identifier. For rare or unseen identifiers, $s_j(id)$ backs off to a $k$-nearest-neighbor estimate over the $K$ closest training identifiers in the feature's vocabulary, where the neighborhood is induced by frequency-weighted token distance.
    \item \textbf{Aggregation.} The example-level score for feature $f_j$ is obtained by aggregating ID-level scores back to the example,
    \begin{equation}\label{eq:agg}
    \hat{y}_i^{(j)} = \frac{1}{k_i} \sum_{t=1}^{k_i} s_j\bigl(id_{i,t}^{(j)}\bigr),
    \end{equation}
    where mean aggregation is used in our experiments. Median and max aggregation produced statistically indistinguishable downstream NE in preliminary studies; mean was selected for stability.
    \item \textbf{Scoring and ranking.} The held-out predictive utility of $f_j$ is computed on $\mathcal{D}^{(j)}_{\text{test,exp}}$ using log loss as the primary criterion, with average precision and AUC retained as secondary diagnostics. Features are ranked in descending order of held-out predictive utility.
\end{enumerate}

The use of large but tractable samples in stage~(1) is a deliberate design choice rather than a stop-gap. Industrial pipelines do not require an exact ranking over the full corpus; they require a stable, actionable ranking obtained quickly enough that the analysis can be rerun whenever the candidate sparse pool changes. Sampling with a held-out split offers a controlled point on this trade-off and is well aligned with how feature owners interact with such pipelines in practice.

\subsection{Theoretical and Computational Properties}\label{subsec:complexity}

The independence assumption underpinning LO-FAR has two consequences. First, the ranking pass scales linearly in the feature count $p$, in contrast to correlation- or interaction-aware methods whose cost typically scales superlinearly in $p$ once pairwise structure is considered. This per-feature marginal screening approach shares the spirit of sure independence screening~\cite{fan2008sure} in classical statistics, adapted here to the sparse ID-list setting. For $n$ training examples, $p$ candidate features, and an average list length $\ell$ per example, the dominant cost per feature is incurred by the construction and lookup of the local ID-level estimator on the exploded data, which gives total complexity
\begin{equation}
\mathcal{O}\bigl(p \cdot n \cdot \ell \cdot \log(n\,\ell)\bigr).
\end{equation}
Second, because each feature is processed in isolation, the workflow exhibits embarrassingly parallel structure across features. A pool of $w$ CPU workers reduces wall-clock time to approximately $\mathcal{O}\bigl(p \cdot n \cdot \ell \cdot \log(n\,\ell) / w\bigr)$ in the limit where worker overhead is negligible. The execution model is CPU-only, model-agnostic, and per-feature independent, which matches commodity production data infrastructure more naturally than repeated GPU retraining loops and supports straightforward caching, retry, and monitoring at per-feature granularity.

\subsection{Hyperparameters and Sensitivity}

LO-FAR exposes a small number of hyperparameters: the sample size $n$, the train/test split ratio, the neighbor count $K$ used in the rare-identifier back-off, and the aggregation function. In our experiments, varying $K$ within a reasonable range had marginal effect on the resulting ranking. Mean, median, and maximum aggregation produced nearly identical downstream NE; mean was selected for stability. Frequent identifiers, which dominate the empirical signal in production logs, are largely insensitive to $K$ because they admit direct estimation from their per-identifier label rate; the back-off rule affects only the rare-identifier tail. These observations are consistent with the expectation that the leading sources of feature-level variance are the vocabulary statistics rather than the precise local model.

\subsection{Scope and Limitations}\label{subsec:scope}

LO-FAR is intentionally a stand-alone, model-agnostic ranking method, and we make its scope explicit in three respects.
\begin{itemize}
    \item \textbf{Interaction-only features:} A feature whose value emerges only through pairwise or higher-order interactions with other features can receive a low LO-FAR score even when its inclusion would benefit the downstream ranker. We position LO-FAR as a first-stage filter rather than a replacement for interaction-aware methods. After first-stage pruning, more expensive procedures can be invoked selectively on the reduced feature pool when the marginal cost is justified.
    \item \textbf{List length:} The current workflow is best suited to short sparse ID-list features, in line with the deployment setting that motivated its design. Longer lists increase the cost of the explosion step in equation~\eqref{eq:explosion} and may require additional approximation, sampling, or sequence-aware estimators. Preliminary post-submission experiments suggest that the workflow continues to produce useful rankings on longer lists at higher cost, but those experiments are outside the scope of the present paper.
    \item \textbf{Ranking as input to retraining:} The output of LO-FAR is a ranking, not a launch decision. Any pruning derived from LO-FAR should be validated through end-to-end retraining of the downstream ranker on the retained subset before deployment. Aggregate scores alone are not sufficient to justify launches in industrial settings.
\end{itemize}

\section{Experimental Setup}\label{sec:experimental_setup}

\subsection{Dataset and Feature Pool}

Experiments are conducted on a production-grade dataset of more than one million logged user--ad interactions, drawn from a large industrial ad recommendation system over an extended observation window in order to span representative variation in user behavior, advertiser inventory, and contextual conditions. After applying standard production filters for data quality and feature eligibility, the candidate sparse feature pool comprises 475 short ID-list features, defined as features whose 99th-percentile list length is below five identifiers. Short lists are the dominant family in the deployed system and are the setting in which LO-FAR was originally designed to operate. The dataset is partitioned into training and held-out evaluation subsets; the training subset is used to fit the LO-FAR local estimators (Section~\ref{sec:local_ranking}) and to retrain the downstream ranker for each method and each feature budget, while the held-out subset is used exclusively for end-to-end NE evaluation.

\subsection{Evaluation Metric: Normalized Entropy}\label{subsec:ne}

We adopt Normalized Entropy (NE) as the primary downstream evaluation metric for both CTR and CVR prediction tasks. This follows standard industrial practice for click-prediction systems \cite{he2014practical}. NE is defined as the average per-example log loss of the model normalized by the entropy of the empirical class distribution,
\begin{equation}
\mathrm{NE} = \frac{\frac{1}{N}\sum_{i=1}^{N} \bigl( y_i \log p_i + (1 - y_i) \log(1 - p_i) \bigr)}{\hat p \log \hat p + (1 - \hat p)\log(1 - \hat p)},
\end{equation}
where $y_i \in \{0, 1\}$ is the binary label, $p_i$ is the model's predicted probability for example $i$, $N$ is the number of evaluation examples, and $\hat p = \frac{1}{N}\sum_i y_i$ is the empirical positive rate. NE is a sensitive and class-imbalance-aware quality measure. Lower NE corresponds to better calibrated predictions, and small absolute reductions translate into commercially meaningful improvements at industrial scale.

We report NE {gain} relative to a fixed reference model. Throughout this paper, the reference model is a dense-only ranker trained on exactly the same logged data as the sparse-augmented models but with all sparse ID-list features removed. The relative NE gain of a method is defined as
\begin{equation}
\Delta\mathrm{NE} = \frac{\mathrm{NE}_{\text{method}} - \mathrm{NE}_{\text{baseline}}}{\mathrm{NE}_{\text{baseline}}},
\end{equation}
where $\mathrm{NE}_{\text{method}}$ is computed on the downstream ranker retrained with the top-$B$ sparse features selected by the candidate ranking method and $\mathrm{NE}_{\text{baseline}}$ is the NE of the dense-only reference. 

\subsection{Baselines}

LO-FAR is compared against three baselines that span the operational spectrum of feature ranking methods used in industrial CTR/CVR pipelines:
\begin{itemize}
    \item \textbf{Naive (coverage-based).} A statistical heuristic that ranks features by simple input properties such as coverage (the proportion of examples in which the feature is non-empty) and basic distributional summaries. The naive baseline has $\mathcal{O}(p)$ cost and is included because, in practice, teams without access to a dedicated ranking workflow frequently rely on similar heuristics. Comparing against it clarifies whether LO-FAR offers material gains over the cheapest plausible alternative.
    \item \textbf{Shuffle-based importance.} A model-agnostic permutation-importance procedure that estimates the importance of each feature by measuring the degradation of model NE when the feature's values are randomly permuted in the held-out set \cite{fisher2019all}. The procedure requires repeated evaluation passes of the downstream model, and stable estimates in industrial CTR/CVR settings typically require recalibration or partial retraining for each feature.
    \item \textbf{Binary Stochastic Neurons (BSN).} A training-coupled feature selection method based on $L_0$-style stochastic gates over input features, jointly optimized with the ranker \cite{louizos2018learningsparseneuralnetworks,yamada2020feature}. BSN is interaction-aware in the sense that gates are learned in the presence of all features simultaneously, and it represents the upper end of the cost--quality trade-off considered here.
\end{itemize}

\subsection{Evaluation Protocol}

For each ranking method $m \in \{\text{Naive}, \text{Shuffle}, \text{BSN}, \text{LO-FAR}\}$ and each feature budget $B \in \{100, 200, 300, 400\}$, we form the top-$B$ subset $\mathcal{F}_B^{(m)}$ produced by $m$, retrain the downstream ranker from scratch on $\mathcal{F}_B^{(m)}$, and report $\Delta\mathrm{NE}$ on the held-out evaluation subset for both CTR and CVR tasks. The choice of budgets covers aggressive ($B=100$), moderate ($B=200$), conservative ($B=300$), and near-full ($B=400$) pruning. As a secondary diagnostic, we report Jaccard similarity over the top-$N$ retained features across pairs of ranking methods. Note that ranking similarity is treated as an interpretability and diagnostic tool, not as a success criterion. The criterion of interest is downstream NE gain at the chosen budget after retraining.

\subsection{Operational Comparison}

Table~\ref{tab:ops} summarizes the operational profile of each method as observed in our environment. LO-FAR completed feature ranking in approximately two CPU-hours when parallelized across CPU workers; the naive baseline completed in minutes. The shuffle-based and BSN procedures required multiple GPU-days when their full retraining and tuning workloads are accumulated. In the same environment and accounting for repeated runs and tuning, the resulting compute spend was below approximately \$100~USD for LO-FAR and exceeded approximately \$4{,}000~USD for each of shuffle-based importance and BSN. We report these numbers as observed operating points in our setting rather than as universal claims; absolute costs depend on hardware, scheduler contention, and the exact tuning regimen used.

\begin{table}[t]
\centering
\footnotesize
\setlength{\tabcolsep}{3pt}
\caption{Operational characteristics of the four ranking workflows. Interact = interaction-aware; Retrain = requires downstream retraining during the ranking pass. Turnaround is as observed in our environment; absolute durations and costs are environment-dependent}
\label{tab:ops}
\begin{tabular}{lccccc}
\toprule
Method & Interact. & Retrain & CPU-only & HW & Turnaround \\
\midrule
Naive    & No       & No  & Yes & CPU & Minutes \\
LO-FAR   & No       & No  & Yes & CPU & $\sim$2 CPU-h \\
Shuffle  & Indirect & Yes & No  & GPU & Multi-day \\
BSN      & Yes      & Yes & No  & GPU & Multi-day \\
\bottomrule
\end{tabular}
\end{table}

The storage implications of the evaluated budgets are deterministic, as discussed in Section~\ref{sec:methodology}. Retaining 100--300 of the 475 sparse ID-list features removes approximately 40--75\% of sparse embedding tables and a comparable fraction of sparse parameters in the deployed model. We emphasize this distinction because, although the storage reduction is not a separate empirical gain, it is a primary driver of the deployment-relevant benefits of pruning. Smaller sparse models lower training-time memory pressure, reduce serving footprint, and shorten retraining cycles. The feature budget therefore acts as a direct systems control rather than only a reporting choice.

\section{Results and Discussion}\label{sec:results}

\subsection{Downstream Quality After Pruning}\label{subsec:downstream}

\begin{figure*}[t]
    \centering
    \subfigure[CTR prediction]{\includegraphics[width=0.49\textwidth]{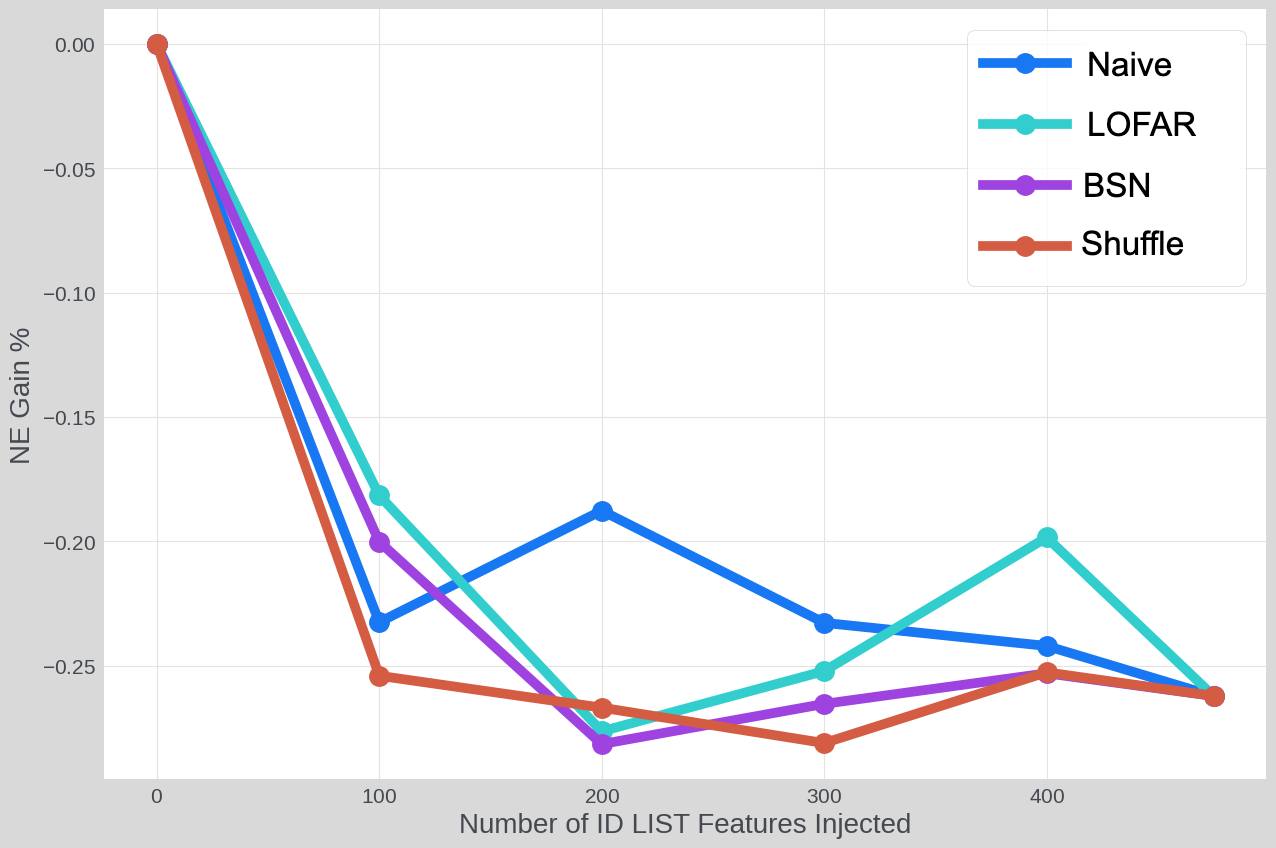}}
    \hfill
    \subfigure[CVR prediction]{\includegraphics[width=0.49\textwidth]{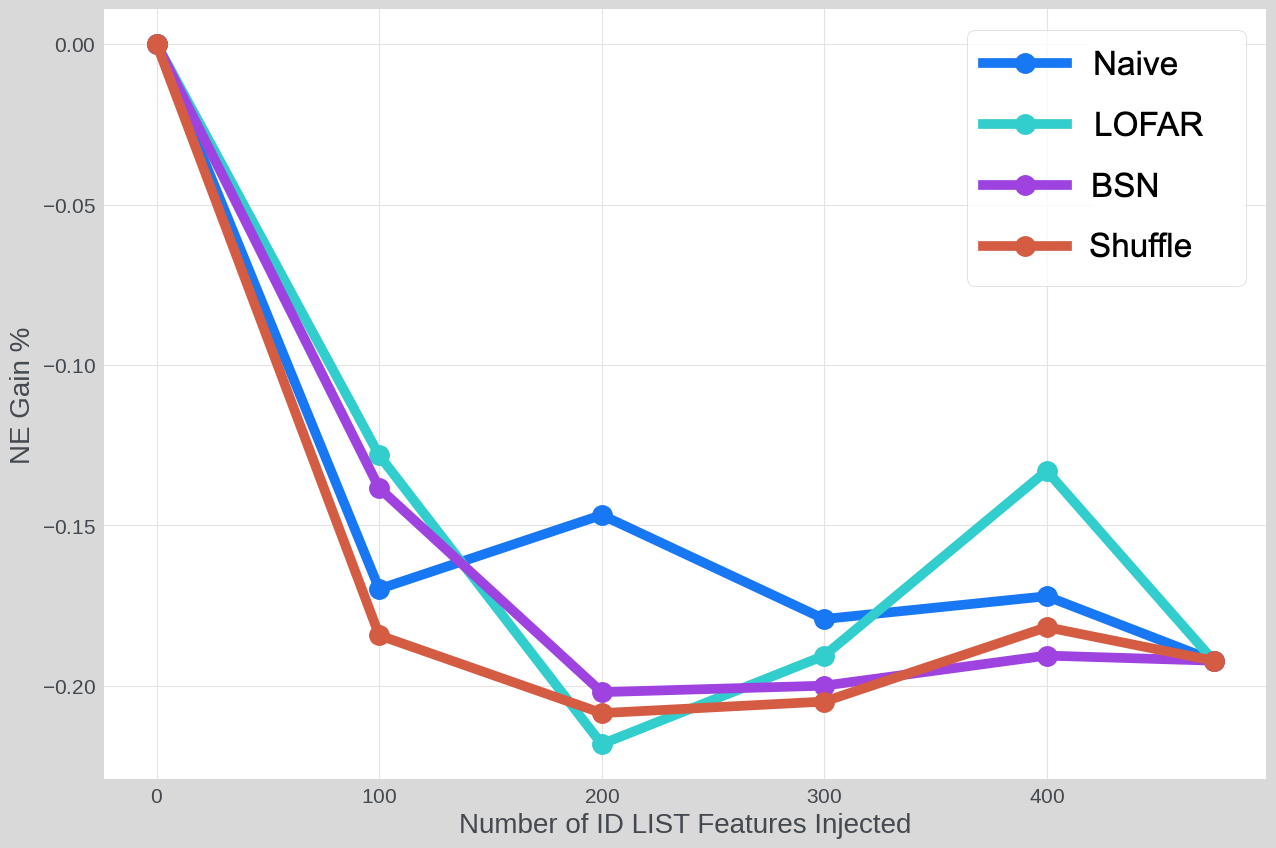}}
    \caption{Downstream NE gain ($\Delta\mathrm{NE}$) relative to a dense-only reference model trained without sparse features. Each curve corresponds to a ranking method, and the horizontal axis sweeps the retained feature budget $B$. LO-FAR remains competitive with shuffle-based importance and BSN across realistic feature budgets while requiring substantially less ranking compute.}
    \label{fig:ne_results}
\end{figure*}

Figure~\ref{fig:ne_results} reports the primary outcome of our evaluation. We plot downstream NE gain $\Delta\mathrm{NE}$ relative to the dense-only reference, measured after retraining the downstream ranker with the top-$B$ sparse features selected by each method. Across both CTR and CVR tasks and across all four feature budgets $B \in \{100, 200, 300, 400\}$, LO-FAR achieves NE gains that are competitive with those obtained by shuffle-based importance and BSN across the evaluated budget range. At individual budget points, the relative ordering among methods varies; the practitioner-relevant finding is that LO-FAR consistently produces a viable quality--cost trade-off across the full range without requiring GPU retraining during the ranking pass. The naive coverage-based baseline tracks LO-FAR closely on the simpler regime of clearly informative features but degrades as the budget tightens. This indicates that LO-FAR captures meaningful held-out predictive structure beyond what coverage statistics alone provide.

The strongest quality--cost trade-off is observed near $B = 200$. At this budget, the retrained ranker preserves a large share of the NE gain available with the full sparse pool while removing more than half of the sparse embedding tables. More conservative operating points ($B = 300$ or $B = 400$) are appropriate when sparse storage is not the dominant constraint and feature owners prefer minimal disruption to the downstream model. More aggressive points ($B = 100$) are appropriate when serving footprint or training memory pressure is the binding constraint. The fact that LO-FAR remains competitive across this range is the practitioner-relevant outcome. It produces a usable quality--cost frontier rather than a single operating point tuned to a particular budget.

\subsection{Ranking Overlap as a Diagnostic}\label{subsec:overlap}

\begin{figure}[t]
    \centering
    \includegraphics[width=1\linewidth]{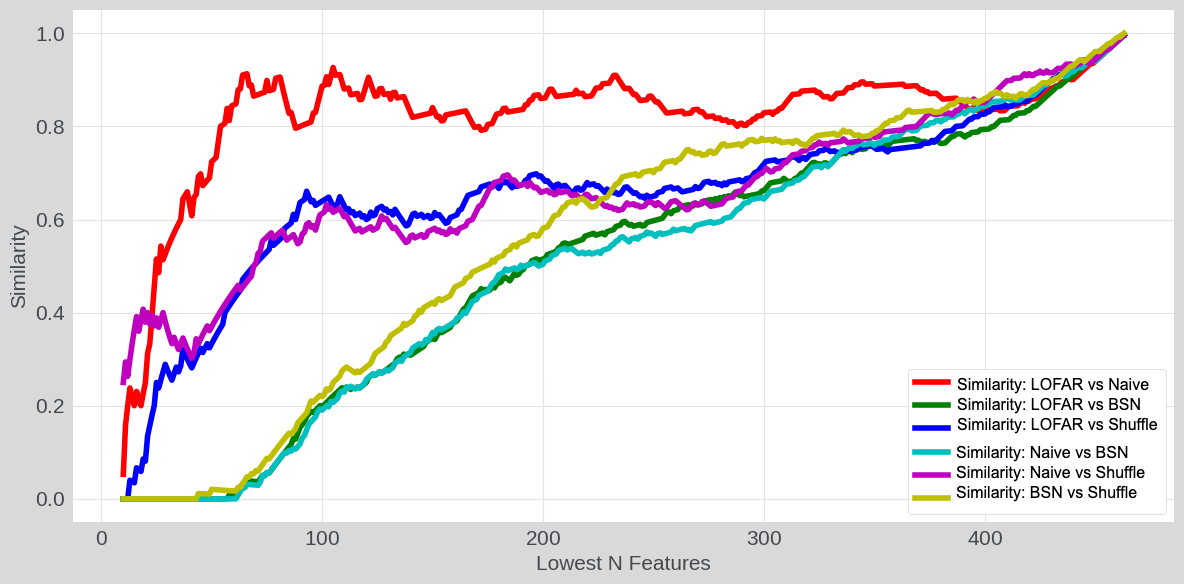}
    \caption{Jaccard similarity over the top-$N$ retained features for each pair of ranking methods. Similarity is reported as a diagnostic only; end-to-end NE gain (Figure~\ref{fig:ne_results}) is the primary evaluation criterion. LO-FAR overlaps strongly with the naive baseline and diverges from BSN, as expected from their differing treatments of feature interactions.}
    \label{fig:topn}
\end{figure}

Figure~\ref{fig:topn} reports Jaccard similarity between the top-$N$ retained features under each pair of ranking methods. The naive heuristic and LO-FAR rankings exhibit relatively high overlap, while both differ markedly from BSN; at $N = 300$, the Jaccard similarity between LO-FAR and BSN is approximately $60\%$. This pattern is expected and follows from the structural difference between the methods. BSN jointly optimizes feature gates in the presence of feature interactions and correlations, and can therefore down-weight features whose marginal value is largely subsumed by other retained features. LO-FAR, by construction, ranks features on stand-alone predictive utility and does not model such redundancy. In high-dimensional sparse spaces, where many features carry overlapping signal, multiple distinct subsets of comparable size can support comparable downstream quality after retraining. Ranking similarity is therefore not a proxy for end-to-end performance.

We retain Figure~\ref{fig:topn} as an interpretability and diagnostic tool, not as a success criterion. The deployment-relevant criterion is the downstream NE gain reported in Section~\ref{subsec:downstream}, which shows that LO-FAR identifies a sufficiently informative subset under each budget despite producing rankings that diverge from BSN.

The overlap pattern also offers a structural explanation for why a local method can be effective at first-stage filtering. Much of the first-stage decision consists of separating clearly weak sparse features from the rest of the pool. Interaction-aware methods are most useful near the difficult boundary where several strong, partially redundant features compete for the final retained slots. In large industrial feature pools, a stand-alone filter such as LO-FAR can therefore remove substantial low-signal capacity at low cost. This leaves the harder near-boundary decisions to more expensive methods if they are warranted.

\subsection{Practical Implications for Industrial Pipelines}\label{subsec:practical}

The results above carry several deployment-relevant implications.

\begin{itemize}

\item{Iteration velocity:} A ranking pass that finishes on commodity CPUs in roughly two CPU-hours is far easier to operate than one requiring repeated GPU retraining. CPU jobs can be scheduled off the accelerator queues used for training and evaluation, removing contention and letting feature owners test new candidate vocabularies or behavioral signals on a near-daily cadence rather than waiting for a centralized retraining slot.

\item{Pipeline integration:} Per-feature independence simplifies operational concerns often invisible in algorithmic comparisons: jobs are easy to parallelize, cache, retry, and monitor; failures are isolated to individual features rather than a global training run; and partial pipelines still yield useful intermediate rankings. These properties matter when ranking is one stage in a larger pipeline managing joins, vocabulary updates, and data snapshots.

\item{Staged use with interaction-aware methods:} The results support a staged workflow rather than a categorical replacement. LO-FAR can shrink a pool of several hundred candidate features into a smaller retained subset, after which interaction-aware methods are applied selectively where their marginal cost is justified---mirroring how many industrial RecSys pipelines reserve expensive procedures for the highest-return parts of the stack.

\item{Organizational consequences:} Because the ranking pass needs no accelerators, it also changes who can run it. Where multiple teams share a ranking platform, feature owners can revisit feature budgets autonomously rather than queuing behind accelerator-bound training and evaluation workloads.
\end{itemize}

\subsection{Competing Interpretations and Boundaries of the Result}\label{subsec:debate}

Two interpretations of the broader literature are consistent with the present findings. The first holds that improvements in ranking architectures \cite{wang2021dcnv2}, distributed training systems \cite{guo2021scalefreectr}, and online embedding infrastructure \cite{liu2022monolith} reduce the urgency of aggressive feature pruning by absorbing more sparse capacity efficiently. The second holds that the same advances increase the value of disciplined sparse-feature management, because each additional feature now enters a more complex and more expensive stack \cite{li2024embeddingcompression,lyu2023optfs}. Our findings are more consistent with the second interpretation in the regime of first-stage filtering on large pools of short ID-list features. We do not claim the broader debate is settled. The evidence reported here is strongest for short ID-list features and budgeted retraining; for longer histories or models with stronger cross-feature dependence, training-coupled selectors may recover gains that justify their additional cost. Identifying the conditions under which each strategy dominates is an open and important question for industry-track research.

\section{Conclusion}\label{sec:conclusion}
We present LO-FAR, a CPU-only local ranking workflow that scores each sparse ID-list feature from its stand-alone held-out predictive signal and decouples first-stage pruning from the downstream training loop. On a production-grade dataset of more than one million logged interactions and 475 candidate features, LO-FAR completes ranking in approximately two CPU-hours, achieves NE gains competitive with shuffle-based importance and BSN across budgets of 100--400 retained features, and yields deterministic 40--75\% reductions in sparse embedding storage. The contribution is system-level rather than purely algorithmic, and argues that feature ranking in industrial recommendation should be evaluated by the joint cost, turnaround time, and operational simplicity of repeated reruns rather than offline expressiveness alone. Under that lens, LO-FAR shifts first-stage sparse feature management toward a repeatable, parallelizable, CPU-bound workflow and complements interaction-aware selectors by leaving them only the harder near-boundary decisions on a smaller, pre-filtered pool. Extending the workflow to longer ID-list features and combining its stand-alone signal with selectively applied interaction-aware steps remain open directions for future work.

\bibliographystyle{ACM-Reference-Format}
\bibliography{references}
\end{document}